\documentclass[prd,preprint,superscriptaddress,preprintnumbers,nofootinbib]{revtex4}
\usepackage{graphicx}
\usepackage{epsfig}
\usepackage{bm}
\usepackage{latexsym,amssymb,amsmath,amssymb,wasysym,float}

\usepackage{color}

\usepackage{enumitem}

\newcommand{\postscript}[2]{\setlength{\epsfxsize}{#2\hsize}
   \centerline{\epsfbox{#1}}}

\usepackage[usenames,dvipsnames]{xcolor}
\definecolor{orange}{cmyk}{0,0.5,1,0}
\definecolor{rossoCP3}{cmyk}{0,.88,.77,.40}
\definecolor{graa}{rgb}{0.8,0.8,0.8}
\definecolor{blaa}{rgb}{0.2,0.2,0.6}

\begin{document}

\title{\color{rossoCP3} Probing
  the Dark Dimension with Auger data}

\author{Neena T. Noble}

\affiliation{Department of Physics and Astronomy,\\ Lehman College, City University of
  New York, NY 10468, USA
}
\affiliation{Department of Astrophysics, American Museum of Natural
  History, NY 10024, USA}

\author{Jorge F. Soriano}

\affiliation{Department of Physics and Astronomy,\\ Lehman College, City University of
  New York, NY 10468, USA
}

\author{Luis A. Anchordoqui}

\affiliation{Department of Physics and Astronomy,\\ Lehman College, City University of
  New York, NY 10468, USA
}

\affiliation{Graduate Center, City University of New York, NY 10016, USA}

\affiliation{Department of Astrophysics, American Museum of Natural
  History, NY 10024, USA}

\begin{abstract}
  \noindent
By combining swampland conjectures with observational data, it was
recently pointed out that our universe could stretch off in an
asymptotic region of the string landscape of vacua. Within this framework, the
cosmological hierarchy problem (i.e. the smallness of the cosmological
constant in Planck units: $\Lambda \sim 10^{-122} M_{\rm Pl}^4$) can
be naturally resolved by the addition of one mesoscopic (dark)
dimension of size $\sim \lambda \, \Lambda^{-1/4} \sim 1~\mu{\rm
  m}$. The Planck scale of the higher dimensional theory, $M_{\rm UV}
\sim  \lambda^{-1/3} \Lambda^{1/12} M_{\rm Pl}^{2/3} \sim 10^{10}~{\rm
  GeV}$, is tantalizingly close to the energy above which the
Telescope Array (TA) and the Pierre Auger collaborations found
conclusive evidence for a sharp cutoff of the flux of
ultra-high-energy cosmic rays (UHECRs). It was recently suggested that since
physics becomes strongly coupled to gravity beyond $M_{\rm UV}$,
universal features deep-rooted in the dark dimension could control the
energy cutoff of the source spectra $\propto E^{-\gamma} \exp(-E/M_{\rm
    UV})$, where $E$ is the cosmic ray energy and $\gamma$ a free
parameter. Conversely, in the absence of phenomena inborn within the dark
dimension, we would expect a high variance of the cosmic ray maximum energy $E_{\rm max}$ characterizing the source spectra $\propto E^{-\gamma}
\exp(-E/E_{\rm max})$, reflecting the many different properties inherent to the most commonly assumed UHECR
accelerators. The most recent analysis of Auger and TA data exposed 
strong evidence for a correlation between UHECRs and nearby starburst
galaxies, with a global significance post-trial of $4.7\sigma$. Since these galaxies are in our cosmic backyard, the flux
attenuation factor due to cosmic ray interactions en route to Earth
turns out to be negligible. This reasoning implies that for each source, the
shape of the observed spectrum should roughly match the emission
spectrum by the starburst, providing a unique testing ground for the dark
dimension hypothesis. Using Auger data, we carry out a maximum
likelihood analysis to characterize the shape of the UHECR emission
from the galaxies dominating the anisotropy signal. We show that 
the observed spectra from these sources could be universal only if $\lambda \alt 10^{-3}$.

\end{abstract}

\maketitle

\section{Introduction}

The origin of ultra-high-energy cosmic rays (UHECRs) is a longstanding
unresolved conundrum in modern astrophysics. Great advances have been
made in uncovering the sources over the past two decades~\cite{Anchordoqui:2018qom}. The most
important is probably the unambiguous detection a flux suppression
near $10^{10.6}~{\rm GeV}$~\cite{HiRes:2007lra,PierreAuger:2008rol}, as predicted by
Greisen, Zatsepin, and Kuzmin
(GZK)~\cite{Greisen:1966jv,Zatsepin:1966jv}. The GZK-limit is an example of the far-reaching connections between
different regimes of physics, tying-up the behavior of the highest
energy particles in nature to the low-energy ($\sim 2.7~{\rm K}$)
photons of the cosmic microwave background (CMB), and can be
explained by sub-GeV scale physics of photo-pion production and/or nucleus
photodisintegration occurring in the extremely boosted relativistic
frame of the cosmic ray. These highly boosted cosmic rays are also incisive probes of fundamental particle-physics properties at
energies far exceeding those at the LHC (for comparison, the
  proton energy of nominal LHC
beam is $7~{\rm TeV}$). In this paper we focus on the particle physics
facet of UHECRs.

While the physics involved in the explanation of the UHECR origin
needs often to be pushed to its extremes to accomodate observations~\cite{Anchordoqui:2018vji,Anchordoqui:2020otc,Anchordoqui:2022pzo},
it is common ground that charged particles could be accelerated in
magnetized astrophysical shocks, whose size and typical magnetic field
strength determine the maximal achievable energy~\cite{Hillas:1984ijl}, akin to the
conditions in (wo)man made particle
accelerators. The most likely astrophysical
accelerators of UHECRs are the shocks associated with:  {\it (i)}~active galactic
nuclei (AGNs)~\cite{Biermann:1987ep}, {\it (ii)}~gamma-ray
bursts (GRBs)~\cite{Waxman:1995vg,Vietri:1995hs}, and {\it (iii)}~powerful superwinds emanating from the
core of starbursting galaxies~\cite{Anchordoqui:1999cu}.

The Pierre Auger Collaboration has reported evidence for a correlation between the arrival directions of the highest energy
cosmic rays and nearby starburst galaxies~\cite{PierreAuger:2018qvk,PierreAuger:2022axr}. When Auger data are
combined with the data set collected by the Telescope Array, there is 
{\it conclusive} evidence for the association, reaching a post-trial significance  of 4.7$\sigma$~\cite{PierreAuger:2023zxh}. Since these galaxies are in our cosmic backyard, the flux
attenuation factor due to cosmic ray interactions en route to Earth
turns out to be negligible. As a matter of fact, the anisotropy signal
coincidentally emerges in the extreme-energy end of the spectrum, above
about the GZK-limit.

Now, in the absence of new physics phenomena, we would expect a high
variance of parameters characterizing the UHECR source spectra beyond
the GZK-limit, reflecting the many different properties inherent to
the acceleration environments in the market. However, the cutoff
spectra could still be universal if some kind of new physics, with a
higher energy loss rate than that of GZK interactions, is
responsible for the cosmic ray maximum energy. In this paper 
we reexamine the test to search for new
physics designed elsewhere~\cite{Anchordoqui:2022ejw}, using the entire data sample collected by the Pierre Auger
Observatory during phase I~\cite{PierreAuger:2022axr}.

The layout of the article is as follows. In Sec.~\ref{sec:2} we review recent
developments in tests of fundamental physics using UHECRs. We
particularized the discussion to the recently proposed dark dimension, 
which provides a natural solution of the cosmological hierarchy
problem~\cite{Montero:2022prj}. In Sec.~\ref{sec:3} we review the Auger
anisotropy signal and provide supplementary compelling arguments underpinning
starburst galaxies as sources of
UHECRs. In Sec.~\ref{sec:4} we
conjecture that starbursts are the sources of UHECRs and study their
individual emission spectra in the search for universality signaling
some new physics phenomena. We particularize the discussion to the
dark dimension and use Auger data to constrain the model. Our
conclusions are collected in Sec.~\ref{sec:5}.

Before proceeding, we pause to note that a recent study of the UHECR spectrum
emerging from a population of sources with a power-law distribution of
maximum energies suggests that source-to-source variance of the
maximum energy must be small to describe the
data~\cite{Ehlert:2022jmy}. This clearly provides further motivation for our study.

\section{Strings at the end of the swampland}
\label{sec:2}

The quickly developing Swampland research program seeks to understand
which are the ``good'' low-energy effective field theories (EFTs) that
can couple to gravity consistently (e.g. the landscape
of superstring theory vacua) and distinguish them from the
``bad'' ones that cannot~\cite{Vafa:2005ui}. In theory space, the border setting apart the
good theories from those relegated to the swampland is characterized
by a set of conjectures on the properties that an EFT should have/avoid
in order to allow a consistent completion into quantum gravity. There
are many swampland conjectures on the block, decidedly too many to be
written down here and so we direct readers to comprehensive reviews~\cite{Palti:2019pca,vanBeest:2021lhn,Agmon:2022thq}.

It has long been known that the smallness of dark energy in Planck units ($\Lambda \sim
10^{-122}M_{\rm Pl}^4$, with $M_{\rm Pl} \sim 1.22 \times 10^{19}~{\rm GeV}$) can be explained statistically~\cite{Bousso:2000xa} or even
anthropically~\cite{Weinberg:1987dv,Susskind:2003kw} by the large number of vacua in the string
landscape. As an alternative, it has been recently proposed~\cite{Montero:2022prj} that when the swampland distance
conjecture~\cite{Ooguri:2006in} is combined with the smallness of dark energy in Planck units (as well
as with other experimental observations on deviations from Newton's gravitational inverse-square law~\cite{Lee:2020zjt} and neutron star heating~\cite{Hannestad:2003yd}), we turn out to be in a peculiar corner of the string
landscape, endowed with a mesoscopic (dark) extra-dimension
characterized by a length-scale in the micron
range.

The swampland distance
conjecture predicts the emergence of infinite towers of states that become
exponentially light~\cite{Ooguri:2006in}. These towers drive a breakdown of the EFT at infinite distance limits
in moduli space. The related anti-de Sitter (AdS) distance conjecture
correlates the dark energy density to the mass scale $m$ of
an infinite tower of states,
\begin{equation}
  m \sim |\Lambda|^\alpha \,,
\end{equation}
  as the negative AdS vacuum energy
 $\Lambda \to 0$, with $\alpha$ a
 positive constant of ${\cal O} (1)$~\cite{Lust:2019zwm}. If we further assume that
 this scaling behavior holds in dS (or
 quasi dS) space (with a positive cosmological constant), the limit $\Lambda \to 0$ also yields an unbounded number of massless modes.

The AdS distance conjecture when generalized to dS space plays a key
role in addressing the cosmological hierarchy problem and associates the 
 length of the dark dimension to the dark energy scale as
 $\Lambda^{-1/4}$, modulo a correction factor $\lambda$. More concretely, the dark dimension opens up at the characteristic mass scale of the
 Kaluza-Klein tower,
\begin{equation}
  m \sim \lambda^{-1} \Lambda^{1/4} \,,
\end{equation}
and physics is described by a 5-dimensional theory up to the
so-called ``species scale''~\cite{Dvali:2007hz,Dvali:2007wp},
\begin{equation}
M_{\rm UV} \sim \lambda^{-1/3} \Lambda^{1/12}
M^{2/3}_{\rm Pl} \,,
\end{equation}
which represents the higher dimensional scale of quantum gravity, with
$10^{-1} < \lambda < 10^{-4}$. 

The dark dimension has the benefits of a rich phenomenology~\cite{Anchordoqui:2022txe,Blumenhagen:2022zzw,Anchordoqui:2022tgp,Gonzalo:2022jac,Anchordoqui:2022svl,Anchordoqui:2023oqm,vandeHeisteeg:2023uxj}. Of particular interest here, in this model the scale of quantum gravity associated
to the higher dimensional theory ($\sim 10^{10}~{\rm GeV}$) is captivatingly close
to the predicted energy of the GZK-limit in the cosmic ray
spectrum~\cite{Montero:2022prj}. In the spirit of~\cite{Anchordoqui:2022ejw}, in what follows we explore whether the observed spectra from nearby sources (i.e. sources
from which cosmic rays would avoid GZK interactions) could be used to
distinguish if the sharp suppression observed in the spectrum is
due to cosmic rays scattering off the CMB or due to new physics
processes inborn within the dark dimension; see the Appendix for an
illustrative example.

\section{Anisotropy from our cosmic backyard}
\label{sec:3}

The Pierre Auger Collaboration reported  evidence for a correlation
between the arrival directions of the highest energy cosmic rays and a
model based on a catalog of bright starburst galaxies~\cite{PierreAuger:2018qvk,PierreAuger:2022axr}. The null
hypothesis of isotropy was tested through an unbinned maximum-likelihood
analysis. The adopted test statistic (TS) for deviation from isotropy being the standard likelihood ratio test
between the starburst-generated UHECR sky model and the null
hypothesis, with
\begin{equation}
    {\rm TS}(\psi,f,E_{\rm min}) = 2 \ln \frac {{\cal L}(\psi,f,E_{\rm
                                   min})}{{\cal L}(\psi,0,E_{\rm min})},
\end{equation}
and likelihood                         \begin{equation}
{\cal L}(\psi,f,E_{\rm min}) = \prod_{E_i \ge
                                 E_{\rm min}} \frac{\Phi(\hat{n}_i;\psi,f) \ \omega(\hat{n}_i)}{\int_{4\pi}
                                 \Phi(\hat{n};\psi,f) \
                                 \omega(\hat{n}) \ d \Omega},
                             \end{equation}
where $\omega(\hat{n})$~is the
                             combined directional exposure of the
                             dataset, $f$ is the anisotropic fraction,
                             and where the UHECR flux model is
                             given by

                             \begin{equation}
    \Phi(\hat{n}; \psi, f) = f \ \Phi_{\rm signal}(\hat{n}; \psi) +
    (1-f) \ \Phi_{\rm background} \, .
  \end{equation}
Here, the contribution of each source is modelled as a von Mises-Fisher
distribution (i.e., the analog of a Gaussian on a 2-sphere) centered on
the source position  
\begin{equation}
    \Phi_{\rm signal}(\hat{n}; \psi) = 
    \frac{1}{\sum_j w_s}
    \sum_j w_s \frac{\psi^{-2}}{4\pi\sinh\psi^{-2}}
    \exp\left(\psi^{-2}\hat{n}_s \cdot \hat{n}\right)
  \end{equation}
  and the background is given by
\begin{equation}
  \Phi_{\rm background} = \frac{1}{4\pi},
\end{equation}
where $E_i$ and $\hat{n}_i$ are the energy and arrival direction of
the $i$-th event, $w_s$ and $\hat{n}_s$ are the weight and position of the $s$-th source candidate,
and $\psi$ is the root-mean-square (rms) deflection per transverse dimension
(i.e. the total r.m.s. deflection
is~$\sqrt{2} \, \psi$ and the equivalent top-hat radius is
  $\Psi = 1.59\psi$).\footnote{Note that the rms of the north-south
    deflection  ($\delta - \delta_0$) is $\psi$ and
the rms of the east-west deflection $\approx (\alpha - \alpha_0) \cos
\delta$ is also $\psi$, so
that the rms of the total deflection is $\sqrt{2} \times \psi $.} For
a given energy threshold, the TS for isotropy follows a $\chi^2$
distribution with two degrees of
freedom.

The analysis is based on the catalog
of~\cite{Lunardini:2019zcf},
which after selected cuts contains 44 starbursts, with distance in the
range $1 \leq d/{\rm Mpc} < 130$. The analysis is repeated by varying
the energy threshold of the selected events between $10^{10.5}~{\rm
  GeV}$ and $10^{10.9}~{\rm GeV}$ in steps of $10^{9}~{\rm GeV}$.

The best-fit model to Auger data yields ${\rm TS} = 25.0$, with
smearing angle $\psi = {15^{+8}_{-4}}^\circ$ and anisotropic fraction
$f = 9^{+6}_{-4}$~\cite{PierreAuger:2022axr}. Remarkably, the energy
threshold of largest statistical significance
$E_{\rm min} \simeq 10^{10.6}~{\rm GeV}$ coincides with the observed
suppression in the spectrum~\cite{HiRes:2007lra,PierreAuger:2008rol},
implying that when we properly account for the barriers to UHECR
propagation in the form of GZK-energy loss
mechanisms~\cite{Greisen:1966jv,Zatsepin:1966jv} we obtain a self
consistent picture for the observed UHECR horizon. The scan in energy
thresholds comes out with a penalty factor, which is estimated through
Monte-Carlo simulations. The post-trial 1-sided Gaussian significance
is $4.0\sigma$. The anisotropy signal is dominated by four galaxies:
NGC 4945, NGC 253, M83, and NGC 1068.\footnote{We note in passing that
  very recently
  the Pierre Auger Collaboration reported the outcome of a
  novel likelihood analysis considering a simultaneous fit of arrival directions,
  energy spectrum, and nuclear composition data~\cite{PierreAuger:2023htc}. The starburst galaxy
  model is favored with a significance of $4.5\sigma$ (taking into account experimental systematic effects) compared to a reference model with only homogeneously distributed background sources.}

Cross-correlation studies with other catalogs have been carried
out, including: {\it (i)}~the large-scale distribution of matter using the Two
Micron All-Sky Survey (2MASS)~\cite{2MASS:2006qir}, taking out sources closer than 1~Mpc; {\it (ii)}~AGNs observed in hard X-rays with {\it
  Swift}-BAT~\cite{Oh:2018wzc}, which includes both radio loud and
quite AGNs; and {\it (iii)}~a selected $\gamma$-AGN sample from the
Fermi-LAT 3FHL catalog~\cite{Fermi-LAT:2017sxy}, which traces jetted
AGNs. For all of these catalogs, the resulting statistical significance is
considerably smaller; namely, ${\rm TS} = 18.0$ for the 2MASS, ${\rm TS} = 19.4$ for
AGNs observed in X-rays, and ${\rm TS} = 17.9$ for jetted-AGNs, in all
casses with similar threshold energy. In addition, we note that given
the ubiquity of GRBs, if these explosions were sources of UHECRs we would expect a
correlation of the highest energy cosmic rays with the 2MASS catalog
as opposed to a particular class of
objects~\cite{Anchordoqui:2019ncn}. Altogether, the preceding discussion clearly
favors starburst galaxies as sources of UHECRs.

Furthermore, when the likelihood analysis is duplicated enlarging the
Auger Phase I data sample with UHECRs detected with the Telescope Array, the
statistical significance of the correlation between the highest energy
cosmic rays and
starburst galaxies increases, reaching a conclusive evidence of $4.7\sigma$~\cite{PierreAuger:2023zxh}.

\section{What can Auger data tell us about UV physics?}
\label{sec:4}

In this section we assume that starburst galaxies are sources of UHECRs and
study the shape of their emission spectra in the search for universality. 

There exists ``lore'' that convinces us that acceleration mechanisms
are driven by the particle's rigidity $R = E/Z$ up to a maximum $R_{\rm
  max}$ leading to consecutive flux suppressions of the elemental
spectra at energies of
\begin{equation}
  E_{\rm max} = Z R_{{\rm max}} \,,
\end{equation} 
where $Z$ is the charge of the UHECR in units of the proton
charge. If the sources of
UHECRs trace such a ``Peters cycle''~\cite{Peters:1961}, emission spectra
characterized by a power law of spectral index $\gamma$ with an
exponential suppression above a cutoff energy $E_{\rm max}$
\begin{equation}
\frac{dN}{dt} \propto E^{-\gamma} \exp\left\{-E/E_{\rm max} \right\}  
\end{equation}
give a good description of the flux and nuclear composition measured at Earth
at ultra-high energies, see e.g.~\cite{PierreAuger:2016use,PierreAuger:2022atd}. However, a major caveat of
this type of analyses is that the sources are typically assumed to be
identical, a description which is highly unlikely to be an accurate
reflection of the many different properties inherent to the acceleration environments in the market. 

Moreover, as we discussed Sec.~\ref{sec:2}, the cutoff spectra could 
be universal (and $Z$ independent) if physics becomes strongly coupled to gravity above
about $10^{10}~{\rm GeV}$. If this were the case, it is reasonable to
assume that the
source spectra would described by
\begin{equation}
\frac{dN}{dt} \propto E^{-\gamma} \exp\left\{-E/E_c \right\} \,,  
\label{spectra}
\end{equation}
where $E_c$ is a critical energy that can be identified, e.g., with
the species scale $M_{\rm UV}$.

\begin{figure}[htb!]
  \begin{minipage}[t]{0.48\textwidth}
    \postscript{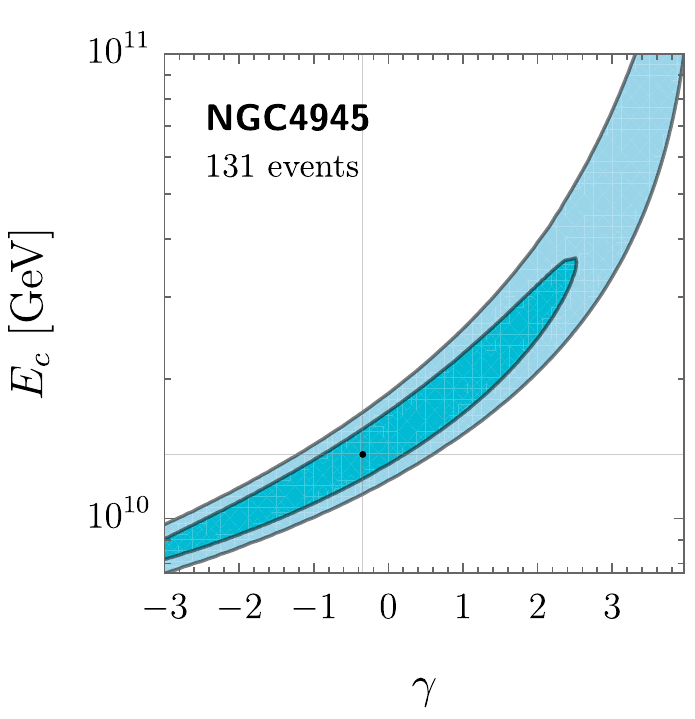}{0.8}
  \end{minipage}
\begin{minipage}[t]{0.48\textwidth}
    \postscript{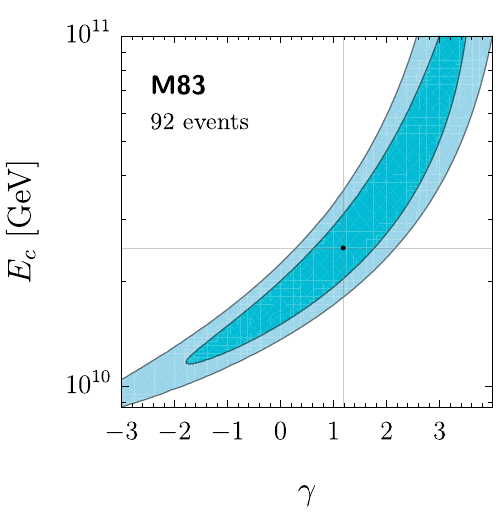}{0.8}
  \end{minipage}\\
  \begin{minipage}[t]{0.48\textwidth}
    \postscript{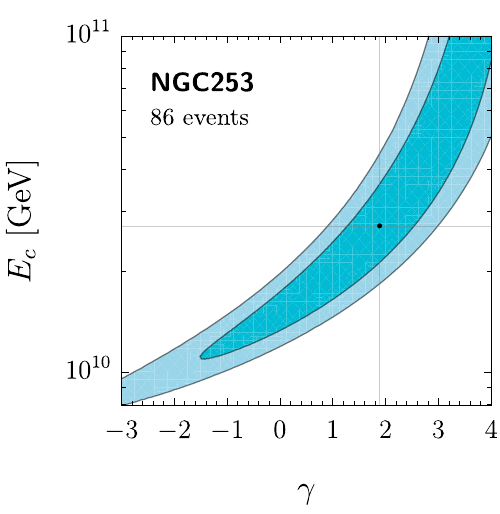}{0.8}
  \end{minipage}
\begin{minipage}[t]{0.48\textwidth}
    \postscript{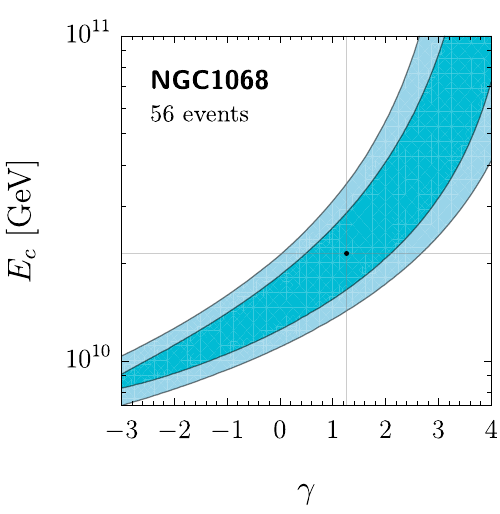}{0.8}
  \end{minipage}\\
\begin{minipage}[t]{0.48\textwidth}
    \postscript{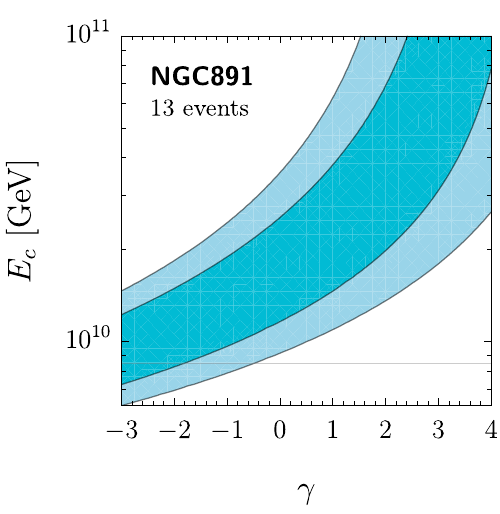}{0.8}
  \end{minipage}
\begin{minipage}[t]{0.48\textwidth}
    \postscript{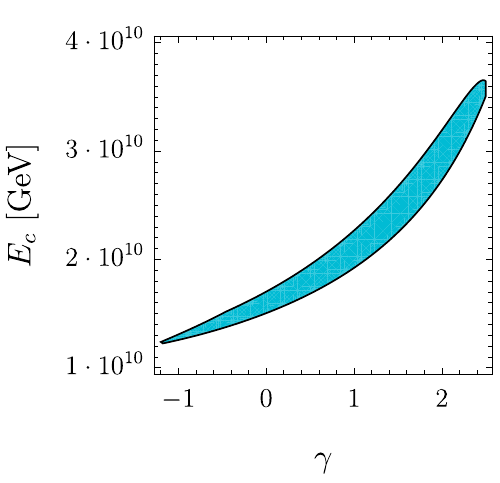}{0.85}
  \end{minipage}
  \caption{68\% and 95\% C.L. contours and the
    overlapping region of the 68\% C.L. contours. \label{figura}}
  \end{figure}

In what follows we performed a likelihood analysis to determine the
parameters $\gamma$ and $E_c$ that best describe the emission
spectrum of each starburst galaxy. In our analysis we consider UHECRs
measured by the Pierre Auger Observatory during Phase I, i.e., 2635
events detected between 2004 and 2020, with energies in the range $3.2
< E/10^{10}~{\rm GeV} < 16.5$~\cite{PierreAuger:2022axr}. The strongest correlation between
UHECRs and starbursts is found to have $E_{\rm min} \simeq 10^{10.6}~{\rm
  GeV}$,  so we ignore events below this energy. We consider the four
starbursts dominating the Auger anisotropy signal and NGC 891, which is
a starburst in a quiescent state~\cite{Anchordoqui:2022pzo}. NGC 891
has a non-negligible contribution to the statistical significance in
the analysis of~\cite{PierreAuger:2023zxh}, but is located outside the
Auger field-of-view. Because of deflections on the Galactic magnetic
field the Auger data sample can contain a few events emitted by NGC
891. The study of this particular source helps to illustrate that the effect on the
search for universality from 
starbursts without significant 
contribution to the Auger anisotropy signal is negligible. For
the selected five sources, we define a circular window around their location
in the sky with angular radius $\theta = 20^\circ$. Such an angular scale
is a good compromise of the favored
top-hat radius $\Psi$ in the Auger
anisotropy search~\cite{PierreAuger:2022axr}. We have
verified that our results do not change significantly by varying the angular
scale within the range $16^\circ <\theta< 24^\circ$. The probability distribution is
given by the normalized spectrum
\begin{equation}
  f(E) = E_c^{\gamma-1} \frac{E^{-\gamma} \ e^{-E/E_c}}{\int_{_{E_{\rm
        th}/E_c}}^{^\infty}  x^{-\gamma} \ e^{-x} \ dx} \,,
  \end{equation}
  with likelihood function
  \begin{equation}
   {\cal L}(\gamma, E_c) \equiv {\rm prob(data|model)} = \prod_{k=1}^N
    f(E_k) \, .
  \end{equation}
Due to the proximity of some of these sources there are events that
could have more than one possible origin.
  
Our results are encapsulated in Fig.~\ref{figura} (where we show the
68\% and 95\% C.L. contours maximizing the likelihood) and
Table~\ref{tabla} (where we give the best fit values of $\gamma$
and $E_c$ together with their 68\% C.L. intervals after
marginalizing over the free parameters). The results can be summarized
as follows:
\begin{itemize}[noitemsep,topsep=0pt]
  \item Using the 68\% C.L. intervals given in Table~\ref{tabla} it is
    straightforward to see that the observed spectra of nearby
    starbursts  can be universal
    in origin if $\gamma \in [-0.3,1.6]$ and $E_c/10^{10}~{\rm GeV}
    \in [1.4,2.4]$.
 \item When the critical energy at the accelerators is identified with the species scale,
   the allowed range of universality leads to the constraint $M_{\rm UV} \agt 10^{10}~{\rm
     GeV}$, corresponding to $\lambda \alt 10^{-3}$.    
\item As can be seen in Fig.~\ref{figura}, the effect of NGC 891
  in constraining the free parameters of the model is negligible.
\item The range of the source spectral index which allows for 
  universality is consistent with
  the one predicted by 
  shock acceleration~\cite{Baerwald:2013pu}.
\end{itemize}

\begin{table}
  \caption{Maximum likelihood estimates and 68\% C.L.
    intervals for $\gamma$ and $E_c$. \label{tabla}}
\begin{tabular}{ccccc}  
\hline
  \hline
 ~~~~Source~~~~ & ~~~~~$\gamma$~~~~~ & ~~~$\gamma$ (68\% C.L.)~~~ &
                                                        ~~~$E_c/10^{10}~{\rm
                                                        GeV}$~~~ &
                                                             ~~~$E_c/10^{10}~{\rm
                                                             GeV}$  (68\% C.L.)~~~ \\
  NGC4945 & $-0.34$ & $[-2.52, 1.61]$ & $1.37$ & $[0.918,2.413]$ \\
  NGC253 & $1.89$ & $[-0.28,3.72]$ & $2.72$ & $[1.403, 10.52]$ \\
  M83 & $1.18$ & $[-0.72,2.83]$ & $2.48$ & $[1.44,6.11]$ \\
  NGC1068 & $1.26$ & $[-1.68,3.69]$ & $2.14$ & $[1.056,9.91]$\\
  NGC891 & $-3.65$ & $[-11.8,2.50]$ & $0.85$ & $[0.391, 4.67]$ \\
  \hline
  \hline
\end{tabular}
\end{table}

\section{Conclusions}
\label{sec:5}

Motivated by the curious case of near-identical cosmic-ray
accelerators~\cite{Ehlert:2022jmy}, we revisited the statistical test
to search for exotic physics  proposed in~\cite{Anchordoqui:2022ejw}
using the entire data sample collected by the Pierre Auger Observatory
during Phase I~\cite{PierreAuger:2022axr}. The test procedure is based
on the search for universality in the UHECR spectra of nearby
sources. UHECRs emitted by these nearby sources can avoid GZK interactions {\it en route} to Earth
and therefore a signal of universality in the spectral shape is a
marker of new physics processes. This is because universality is at odds with the typically high variances of intrinsic properties for the most commonly assumed astrophysical sources.

The study was performed
through a maximum likelihood method. The dramatic
growth in the number of UHECRs when compared to the data sample
considered in~\cite{Anchordoqui:2022ejw} allowed us to refine the
functional form of the probability distribution, leaving two parameters
free in the likelihood analysis: the source spectral index $\gamma$ and the
cutoff energy $E_c$, herein identified with the species scale $M_{\rm UV}$. We have shown
that the spectra of nearby sources can be universal if $M_{\rm UV}
\agt 10^{10}~{\rm GeV}$, corresponding to $\lambda \alt 10^{-3}$.

\section*{Acknowledgments}

We have benefitted from discussions with our colleagues of the Pierre
Auger Collaboration. L.A.A. is supported by the U.S. National Science
Foundation (NSF) Grant PHY-2112527. N.T.N. is supported by the
AstroCom NYC program through NSF Grant AST-2219090. J.F.S. is
supported by Schmidt Futures, a philanthropic initiative founded by
Eric and Wendy Schmidt, as part of the Virtual Institute for
Astrophysics (VIA).

\section*{Appendix}

It has long been known that charged particles moving above a
diffraction grating (without
crossing it) emit Smith–Purcell (SP) electromagnetic radiation~\cite{Smith:1953sq}. As most radiation processes
accompanying the motion of charged particles, the SP effect can be
explained as a result of scattering of the Coulomb field of the moving charged
particles on the irregularities of media. It is important to note that
the charged particles are not scattered in the material of the
target. For this reason, the flow of charged particles does not damage
the target, providing the reliability and long survival time for
practical SP based radiation sources.

The gravitational analog of the SP effect was first entertained
in~\cite{Cardoso:2006nz} as a mechanism that could degrade the energy of UHECRs; the role of the
difraction grating being played by the inhomogeneities of the extra-dimensional space, such as
a hidden brane at finite distance.

\begin{figure}[htb!]
    \postscript{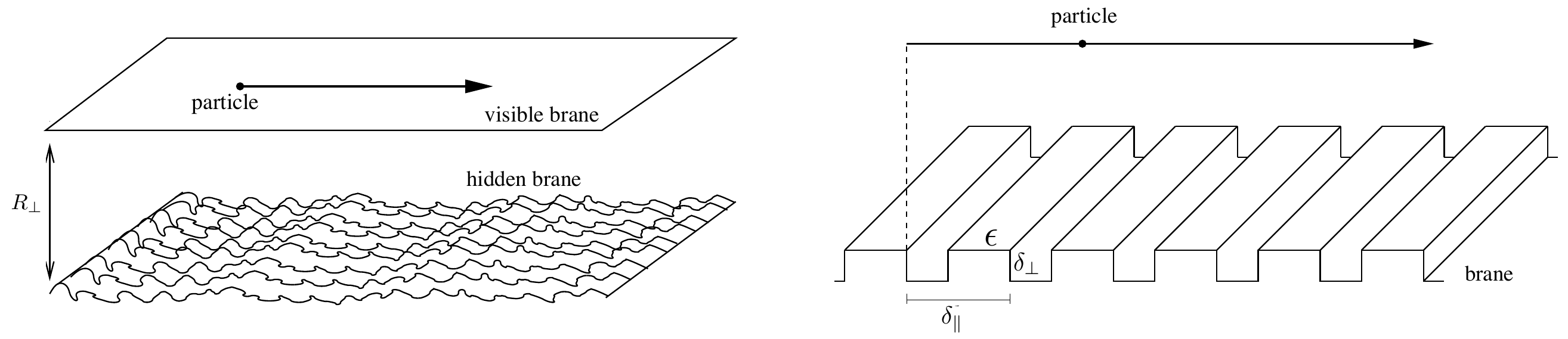}{0.99}
\caption{Pictorial representation of the scenario described in the
  text (left) with periodic longitudinal brane perturbations $\epsilon$ (right). \label{fig:cartoon}}
\end{figure}

Envision the situation protrayed in
Fig.~\ref{fig:cartoon}, in which an UHECR moves parallel and close to an inhomogeneous hidden brane. A periodically
inhomogeneous surface is just one of the special cases of a
grating. Thus, the hidden brane can be 
visualized here as a diffraction
grating, with the cosmic ray propagating in a direction perpendicular to the grating
rulings. To generate gravitational radiation the cosmic ray must
transfer momentum along the particle's trajectory to the inhomogeneous
structure.

\begin{figure}[htpb!]
    \postscript{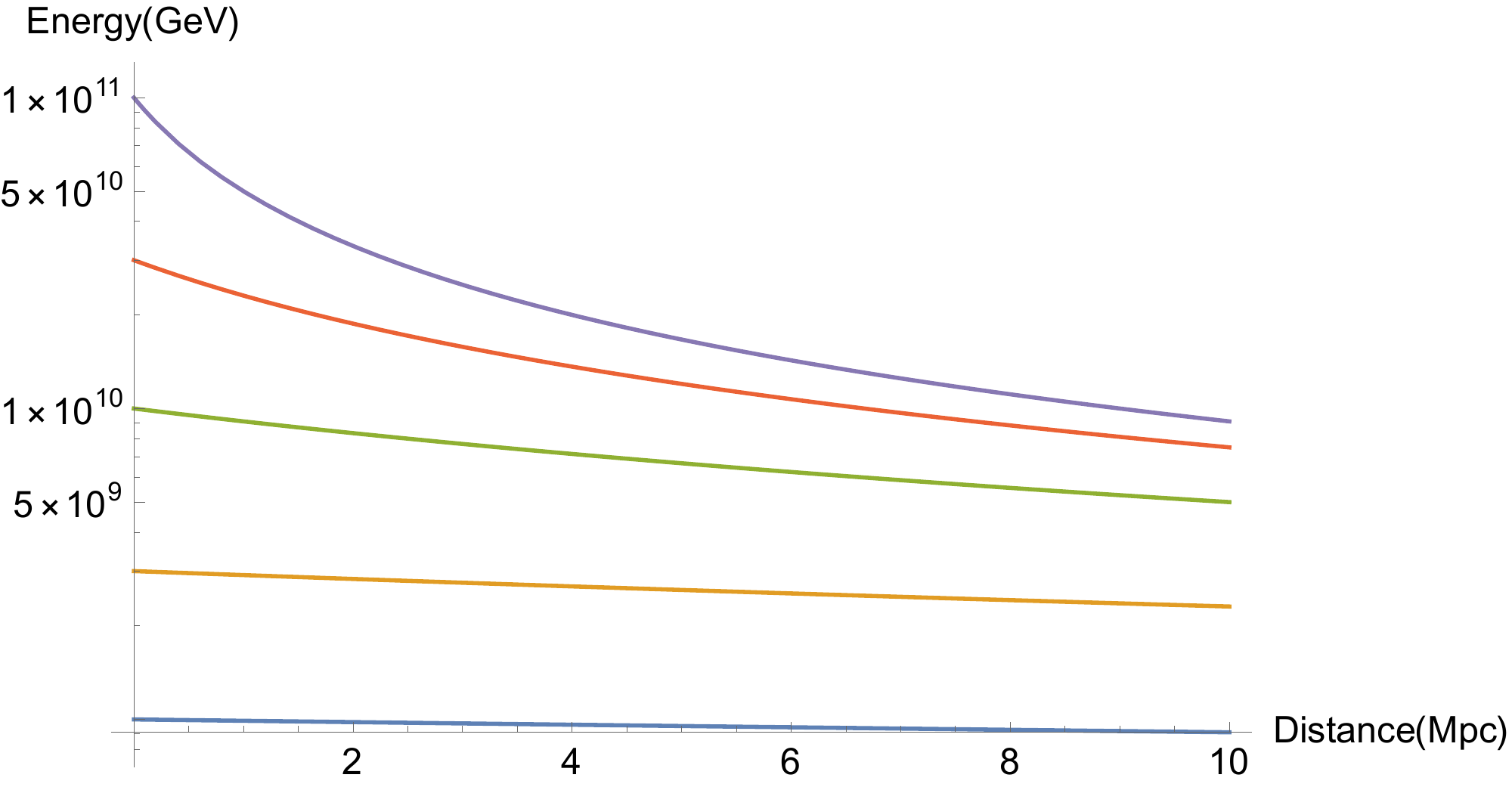}{0.9}
\caption{Energy attenuation length of cosmic rays in the intergalactic medium. \label{fig:E}}
\end{figure}

As an illustration of this gravitational phenomenon, we compute the
power radiated by UHECRs in the presence of a hidden brane, with
typical longitudinal perturbations of length scale $\epsilon$ and transverse perturbations of length scale $\delta_\perp$. Following~\cite{Cardoso:2006nz}, we model these perturbations as a $\delta_\parallel$-periodic
lamellar grating with rulings of width
$\epsilon$ perpendicular to the particle direction of motion; see
Fig.~\ref{fig:cartoon}. The energy loss per unit distance due to graviton emission is
estiamted to be
\begin{eqnarray}
  \frac{d\ln E}{dx} & \sim & -8\pi \frac{E}{M_{\rm UV}^3} \
  \frac{\delta_\perp^2}{\delta_\parallel^5} \ \exp\left\{-\frac{2 \pi
  R_\perp}{\Gamma \delta_\parallel} \right\} \nonumber \\ 
  & \sim & - 0.1 \left(\frac{E}{10^{10}~{\rm GeV}}\right)~{\rm Mpc}^{-1}
           \, , 
\end{eqnarray}           
where $\Gamma$ is the UHECR Lorentz boost, $R_\perp \sim 1/m$ the
compactification radius, $\delta_\parallel = 2
\epsilon$, and where in the second rendition we have taken
$\delta_\perp \sim 0.1 \delta_\parallel$ and $\delta_\parallel \sim  10^{-3} R_\perp$~\cite{Anchordoqui:2022ejw}. Note
that for $\Gamma > 10^{5}$ the fractional energy loss traces the
particle's energy and is independent of  the Lorentz boost. In
Fig.~\ref{fig:E} we show the energy
attenuation length of cosmic rays in the intergalactic medium. Due to the fast energy loss rate above the
species scale $M_{\rm UV}$  it is reasonable to assume the source
spectra can be described by (\ref{spectra}).

\end{document}